\documentclass{sigchi-ext}

\copyrightinfo{Presented at the workshop "The Future of Hybrid Care and Wellbeing in HCI", April 23, 2023, Hamburg, Germany}

\usepackage[english]{babel}

\usepackage[T1]{fontenc}
\usepackage{textcomp}
\usepackage[scaled=.92]{helvet} % for proper fonts
\usepackage{graphicx} % for EPS use the graphics package instead
\usepackage{balance}  % for useful for balancing the last columns
\usepackage{booktabs} % for pretty table rules
\usepackage{ccicons}  % for Creative Commons citation icons
\usepackage{ragged2e} % for tighter hyphenation; defines new commands \Centering, \RaggedLeft, and \RaggedRight
\usepackage{dirtytalk} % typeset quotations
\usepackage[utf8]{inputenc}  % translates  standard and other input encodings into a ‘LATEX internal language’
\usepackage{todonotes} % user mark things to do later, in a simple and visually appealing way
\usepackage{float} % Improves the interface for defining floating objects such as figures and tables.
\usepackage{subcaption} % captions for subfigures and the like
\usepackage{txfonts}
\usepackage{url}
\makeatletter
\g@addto@macro{\UrlBreaks}{\UrlOrds}
\makeatother

\usepackage{multirow} %tabular cells spanning multiple rows
\usepackage{tabularx} %defines an environment tabularx, an extension of tabular which has an additional column designator, X, which creates a paragraph-like column whose width automatically expands so that the declared width of the environment is filled

\usepackage{color} %colour management

\usepackage{enumitem} %better  enumerate, itemize and description

\usepackage[official]{eurosym}
\usepackage{siunitx} % Typesetting values with units

\usepackage[normalem]{ulem}%Package for underlining in references

\usepackage{microtype}        % Improved Tracking and Kerning
\usepackage[all]{hypcap}      % Fixes bug in hyperref caption linking
\usepackage{ccicons}          % Cite your images correctly!

\usepackage{cite} %makes citations in a block in correct order (not possible in 'Manuscript')

\def\plaintitle{Discussing Risks and Benefits in the Future of Hybrid Rehabilitation and Fitness in Mixed Reality} 
  \def\plainauthor{Jana Franceska Funke, Enrico Rukzio}

\def\plainkeywords{hybrid care, hybrid workout, hybrid rehabilitation, virtual reality}

\title{\plaintitle}

\numberofauthors{2}
\author{
  \alignauthor{%
    \textbf{Jana Funke}\\
    \affaddr{Institute of Media Informatics} \\
    \affaddr{Ulm University, Germany} \\
    \email{jana.funke@uni-ulm.de} } \vfil 
    \alignauthor{%
    \textbf{Enrico Rukzio}\\    
    \affaddr{Institute of Media Informatics} \\
    \affaddr{Ulm University, Germany} \\
    \email{enrico.rukzio@uni-ulm.de} }
    }

\definecolor{linkColor}{RGB}{6,125,233}
\hypersetup{%
  pdftitle={\plaintitle},
  pdfauthor={\plainauthor},
  pdfkeywords={\plainkeywords},
  bookmarksnumbered,
  pdfstartview={FitH},
  colorlinks,
  citecolor=black,
  filecolor=black,
  linkcolor=black,
  urlcolor=linkColor,
  breaklinks=true,
}

\begin{document}

%% For the camera ready, use the commands provided by the ACM in the Permission Release Form.
%\CopyrightYear{2007}
%\setcopyright{rightsretained}
%\conferenceinfo{WOODSTOCK}{'97 El Paso, Texas USA}
%\isbn{0-12345-67-8/90/01}
%\doi{http://dx.doi.org/10.1145/2858036.2858119}       
%% Then override the default copyright message with the \acmcopyright command.
%\copyrightinfo{\acmcopyright}

\maketitle

% Uncomment to disable hyphenation (not recommended)
% https://twitter.com/anjirokhan/status/546046683331973120
\RaggedRight{} 

% Do not change the page size or page settings.
\begin{abstract}
In a world where in-person context transitions more into remote and hybrid concepts, we should consider new concepts of interaction in health and rehabilitation and what advantages and disadvantages they bring. 
One of the rising topics is mixed reality, where we can use the advantages of immersive 3D, 360-degree environments. 
Meanwhile, physical activity is further decreasing and with it negative effects increase through sedentary behaviour or wrong and untrained movements. 
In this position paper, we discuss these new risks and potential benefits of mixed reality technology when used for rehabilitation and fitness.
We conclude with suggesting better feedback and guidance for physical movement and tasks at home. 
Improving feedback and guidance for participants could be achieved through using new technologies like virtual reality and motion tracking.  
\end{abstract}

\keywords{\plainkeywords}

\begin{CCSXML}
<ccs2012>
<concept>
<concept_id>10003120.10003121.10003126</concept_id>
<concept_desc>Human-centered computing~HCI theory, concepts and models</concept_desc>
<concept_significance>300</concept_significance>
</concept>
</ccs2012>
\end{CCSXML}

\ccsdesc[500]{Human-centered computing~HCI theory}
\printccsdesc

\newpage
\section{Introduction} % 1 Seite
%1. Motivation: What is the real-world problem that we are trying to solve?

Tomorrow's healthcare will change significantly as a result of digitization, and in many areas, telehealth concepts will enable a spatial and temporal separation between medical staff and patients. 
Telehealth allows for more flexible, faster, sustainable location-independent as well as potentially more cost-effective screening, detection, and treatment \cite{bhatInfrastructuringTelehealthFormal2021,mentisCraftingImageSurgical2016,hongSupportingFamiliesReviewing2017,kocaballiPersonalizationConversationalAgents2019}. 
Hybrid solutions are growing with a more and more connected world, driven by the idea of overcoming distances with virtual platforms. % to stay connected and closer to each other. 
This trend was heavily fueled by the COVID-19 pandemic, where lockdowns made even very close distances in the real world impossible to reach \cite{aksoyWorkingHomeWorld2022}.
Looking into the future, the concept of hybrid connectivity will be important for sustainable collaboration at close and far distances. 

%2. Problem: Why is it important to solve this problem
At the same time, we currently observe a steady trend towards physical inactivity or sedentary behavior \cite{dunstanTooMuchSitting2012a} and obesity \cite{mansonEscalatingPandemicsObesity2004}. 
Although health research points out the many related dangers like diabetes, cardiovascular risk, bad stress regulation\cite{enselPhysicalFitnessStress2004} and mental health\cite{rodriguez-ayllonRolePhysicalActivity2019,gauvinPhysicalActivityPsychological1996} of constant sitting or being overweight.
For this reason, suitable digital and distributed formats for self and professional health care are becoming increasingly important. 
This also applies to physical and movement therapy, which should ideally also be able to be continued or carried out at everyone's home. 
There are now a variety of offerings that use mobile apps\cite{BoostYourInteractive}, online videos \cite{OnlineShoulderRehab}, or video conferencing-based solutions \cite{PhysicalTherapyVirtual} to provide sports and fitness classes, physical therapy sessions, and other exercise training. 
The vast majority of these solutions are presented on a two-dimensional screen (smartphone, tablet, TV). 
However, this often means that complex movement sequences can only be displayed to a limited extent: occlusion or screen size, lowering three-dimensional understanding of the movement, personal feedback from a therapist or trainer, and social get-together compared to real-life opportunities.   

%3. Solution: What is the solution that we came up with to solve it?
Immersive Extended Reality (XR) solutions like virtual reality (VR) or augmented reality (AR) can overcome some of these challenges by connecting people through time and distance in the same virtual space.
Furthermore, three-dimensional solutions can extend real-life feedback through fast, changeable on-body or in-the-room markings, visualization of movement analysis, and physiological data. 
However, feedback and visualizations have to be helpful for both the patient and physiotherapist, supporting physical movements with minimal negative side effects.
Mixed reality approaches for physical activity are very promising and will likely be more accepted in the near future.
The market development for home fitness is growing steadily and is expected to increase by 33\% from 2019 to 2027. The market is expected to be worth \$59,231 million dollars by then \cite{OnlineFitnessMarket}. 
Since mixed reality will be a very large market and important in the future\cite{ARVRMarket}, home fitness and rehabilitation in combination with mixed reality devices is likely to be a widespread option.

\section{Discussing Benefits in Remote Telephysiotherapy and Teletraining}
Supporting physical activity as Telephysiotherapy and Teletraining with mixed reality headsets can have many benefits. 

\paragraph{Real Time Visualization}
VR and AR have the potential to manipulate the environment and virtual objects in real time. 
This means training, tasks, guidance, and feedback on mistakes can be visualized while training to avoid getting into the habit of incorrect movements. 
Furthermore, body parts can easily be marked, highlighted, or manipulated to support learning, training, and feedback in a user-friendly and customized way. 
Some literature is already trying to use AR for human motion analysis \cite{debarbaAugmentedRealityVisualization2018a} by visualizing the joint range of motion. Other approaches try to support posture guidance with visual and haptic feedback \cite{schonauerMultimodalMotionGuidance2012b}.
This can partly replace or improve workout tasks that have to be done without a therapist or a trainer. When therapists have to rely on the exercises being performed correctly, they can monitor the workout, and patients get immediate feedback if the tasks are not executed correctly.

\paragraph{No Travel Distances}
Instead of going to a physiotherapist's office, fitness studio, or sports place, users don't have to overcome the distance. 
They can stay at home or wherever they are right now. 
Dropping out the distance that has to be covered saves time that would have been spent most likely getting to the therapist's office. 
Without driving, whether by car or public transport, capacities on traffic routes can be saved, which promotes CO2 reduction and environmental protection.

\paragraph{Worldwide Connection}
Distances that can not be easily overcome, e.g., other countries or continents, can meet in the virtual environment. Therapists can have clients everywhere and can support them from everywhere. Furthermore, trainees, friends or patients can be connected, meet, and work in a group to motivate each other. 

\paragraph{Motivation}
Concerning motivation, we find many new possibilities for XR to take advantage of. We can not only manipulate our own body to improve performance and motivation\cite{kocurFlexingMusclesVirtual2020a}. We can change the environment\cite{thompsoncoonDoesParticipatingPhysical2011} and the shape of people training together with us in the XR world.

\section{Discussing Risks in Remote Telephysiotherapy or Teletraining}

While supporting XR physical activity (as Telephysiotherapy and Teletraining) has elicited advantages, it can further have it harbours risks that must be taken into account.  

\paragraph{Stronger and Worldwide Competition}
Through technical support and worldwide connection, distance from a trainer or therapist is no longer a barrier. Having access to all online available therapists, results in stronger competition and possibly reduces changes to establish new therapists or trainers. A therapist must not only be the best in town to get new clients but the best "in the world". The only limitation will be the time the therapist has, but with further technical support, the therapist has to spend less time on patients. 

\paragraph{Bubble Effect}
The worldwide connection can further support already happening ``bubble effects'' \cite{spohrFakeNewsIdeological2017}. 
When participants can choose where, when, and with whom to work out, from whom to be advised and treated, and with which digital system of choice, bubble effects will appear.
While at the moment, physical places like fitness studios and therapists' offices are mostly chosen by distance relevant choices. 
Different people from different social and cultural environments, with different opinions, inevitably meet each other. However, with more digital offerings, confrontation with people, trainers or therapists with different opinions or training styles will further decrease.

\paragraph{Possible Social Impoverishment for Single Living Persons}
On the other side, for people living alone, digital approaches from home can result in social impoverishment because they are not forced to meet other people in real life. 
Especially for older alone living people, loneliness can affect physical and mental health \cite{jamesalexandercrewdsonEffectLonelinessElderly2016}. 
Physical contact is enormously important for health and can be somewhat replaced by nursery \cite{donaldsonLonelinessElderlyPeople1996} or even by touch in a therapy session \cite{heatleytejadaPhysicalContactLoneliness2020}. 
These negative effects of social isolation happened during the COVID-19 pandemic and need to be avoided in the future \cite{sayinkasarLifeLockdownSocial2021,nooneLonelyTouchNarrative2022}.

\paragraph{Effects of Home Isolation}
Apart from the loneliness factor of living alone people, the development of not leaving the house because everything can be done from home can have far-reaching consequences. These effects range from well-being\cite{issaSocialIsolationPsychological2021}, depression \cite{taylorSocialIsolationDepression2018}, low non-activity related physical activity \cite{levineNonexerciseActivityThermogenesis2002} to functional decline \cite{fujiwaraSynergisticIndependentImpacts2017}. 

\section{Conclusion}
A good balance between in-person and digital health seems to be essential for a satisfied and healthy society. 
It has to be well considered when to resort to digital support and when to aim for in-person solutions, considering all factors from sustainability, urgency, time and location constraints to social factors such as loneliness. 
Therefore, hybrid approaches should be more important than only digital or in-person approaches.
 I would be delighted to discuss these topics further in the workshop with participants from different backgrounds who can give new/other/complementary perspectives on these topics. Further, I would like to discuss the topics I am working on and elaborate on new approaches together.

\bibliographystyle{SIGCHI-Reference-Format}
\bibliography{biblio.bib}


\begin{thebibliography}{00}

%%% the Web address.

\ifx \showCODEN    \undefined \def \showCODEN     #1{\unskip}     \fi
\ifx \showDOI      \undefined \def \showDOI       #1{{\tt DOI:}\penalty0{#1}\ } \fi
\ifx \showISBNx    \undefined \def \showISBNx     #1{\unskip}     \fi
\ifx \showISBNxiii \undefined \def \showISBNxiii  #1{\unskip}     \fi
\ifx \showISSN     \undefined \def \showISSN      #1{\unskip}     \fi
\ifx \showLCCN     \undefined \def \showLCCN      #1{\unskip}     \fi
\ifx \shownote     \undefined \def \shownote      #1{#1}          \fi
\ifx \showarticletitle \undefined \def \showarticletitle #1{#1}   \fi
\ifx \showURL      \undefined \def \showURL       #1{#1}          \fi

\bibitem{aksoyWorkingHomeWorld2022}
{Cevat~Giray Aksoy}, {Jose~Maria Barrero}, {Nicholas Bloom}, {Steven~J. Davis}, {Mathias Dolls}, {and} {Pablo Zarate}. 2022.
\newblock Working from {{Home Around}} the {{World}}.
\newblock   (Sept. 2022).
\newblock
\showDOI{%
\url{http://dx.doi.org/10.3386/w30446}}


\bibitem{bhatInfrastructuringTelehealthFormal2021}
{Karthik~S Bhat}, {Mohit Jain}, {and} {Neha Kumar}. 2021.
\newblock \showarticletitle{Infrastructuring {{Telehealth}} in ({{In}}){{Formal Patient-Doctor Contexts}}}.
\newblock {\em Proceedings of the ACM on Human-Computer Interaction\/} {5}, CSCW2 (Oct. 2021), 323:1--323:28.
\newblock
\showDOI{%
\url{http://dx.doi.org/10.1145/3476064}}


\bibitem{BoostYourInteractive}
{Boost}. 2020.
\newblock Boost - {{Your}} Interactive Personal {{Trainer}}.
\newblock https://www.boost.fit/.   (2020).
\newblock
\newblock
\shownote{(Accessed 13.06.2022).}


\bibitem{debarbaAugmentedRealityVisualization2018a}
{Henrique~Galvan Debarba}, {Marcelo~Elias {de Oliveira}}, {Alexandre L{\"a}dermann}, {Sylvain Chagu{\'e}}, {and} {Caecilia Charbonnier}. 2018.
\newblock \showarticletitle{Augmented {{Reality Visualization}} of {{Joint Movements}} for {{Physical Examination}} and {{Rehabilitation}}}. In {\em 2018 {{IEEE Conference}} on {{Virtual Reality}} and {{3D User Interfaces}} ({{VR}})}. 537--538.
\newblock
\showDOI{%
\url{http://dx.doi.org/10.1109/VR.2018.8446368}}


\bibitem{donaldsonLonelinessElderlyPeople1996}
{Jean~M Donaldson} {and} {Roger Watson}. 1996.
\newblock \showarticletitle{Loneliness in Elderly People: An Important Area for Nursing Research}.
\newblock {\em Journal of Advanced Nursing\/} {24}, 5 (1996), 952--959.
\newblock
\showISSN{1365-2648}
\showDOI{%
\url{http://dx.doi.org/10.1111/j.1365-2648.1996.tb02931.x}}


\bibitem{dunstanTooMuchSitting2012a}
{David~W. Dunstan}, {Bethany Howard}, {Genevieve~N. Healy}, {and} {Neville Owen}. 2012.
\newblock \showarticletitle{Too Much Sitting \textendash{} {{A}} Health Hazard}.
\newblock {\em Diabetes Research and Clinical Practice\/} {97}, 3 (Sept. 2012), 368--376.
\newblock
\showISSN{0168-8227}
\showDOI{%
\url{http://dx.doi.org/10.1016/j.diabres.2012.05.020}}


\bibitem{enselPhysicalFitnessStress2004}
{Walter~M. Ensel} {and} {Nan Lin}. 2004.
\newblock \showarticletitle{Physical Fitness and the Stress Process}.
\newblock {\em Journal of Community Psychology\/} {32}, 1 (2004), 81--101.
\newblock
\showISSN{1520-6629}
\showDOI{%
\url{http://dx.doi.org/10.1002/jcop.10079}}


\bibitem{fujiwaraSynergisticIndependentImpacts2017}
{Yoshinori Fujiwara}, {Mariko Nishi}, {Taro Fukaya}, {Masami Hasebe}, {Kumiko Nonaka}, {Takashi Koike}, {Hiroyuki Suzuki}, {Yoh Murayama}, {Masashige Saito}, {and} {Erika Kobayashi}. 2017.
\newblock \showarticletitle{Synergistic or Independent Impacts of Low Frequency of Going Outside the Home and Social Isolation on Functional Decline: {{A}} 4-Year Prospective Study of Urban {{Japanese}} Older Adults}.
\newblock {\em Geriatrics \& Gerontology International\/} {17}, 3 (2017), 500--508.
\newblock
\showISSN{1447-0594}
\showDOI{%
\url{http://dx.doi.org/10.1111/ggi.12731}}


\bibitem{gauvinPhysicalActivityPsychological1996}
{L. Gauvin} {and} {J.~C. Spence}. 1996.
\newblock \showarticletitle{Physical Activity and Psychological Well-Being: Knowledge Base, Current Issues, and Caveats}.
\newblock {\em Nutrition Reviews\/} {54}, 4 Pt 2 (April 1996), S53--65.
\newblock
\showISSN{0029-6643}
\showDOI{%
\url{http://dx.doi.org/10.1111/j.1753-4887.1996.tb03899.x}}


\bibitem{heatleytejadaPhysicalContactLoneliness2020}
{A. Heatley~Tejada}, {R.~I.~M. Dunbar}, {and} {M. Montero}. 2020.
\newblock \showarticletitle{Physical {{Contact}} and {{Loneliness}}: {{Being Touched Reduces Perceptions}} of {{Loneliness}}}.
\newblock {\em Adaptive Human Behavior and Physiology\/} {6}, 3 (Sept. 2020), 292--306.
\newblock
\showISSN{2198-7335}
\showDOI{%
\url{http://dx.doi.org/10.1007/s40750-020-00138-0}}


\bibitem{hongSupportingFamiliesReviewing2017}
{Matthew~K. Hong}, {Clayton Feustel}, {Meeshu Agnihotri}, {Max Silverman}, {Stephen~F. Simoneaux}, {and} {Lauren Wilcox}. 2017.
\newblock \showarticletitle{Supporting {{Families}} in {{Reviewing}} and {{Communicating}} about {{Radiology Imaging Studies}}}.
\newblock {\em Proceedings of the SIGCHI conference on human factors in computing systems. CHI Conference\/}  {2017} (May 2017), 5245--5256.
\newblock
\showDOI{%
\url{http://dx.doi.org/10.1145/3025453.3025754}}


\bibitem{ARVRMarket}
{Statista. Inc}. 2021.
\newblock {{AR}}/{{VR}} Market Size Worldwide 2021-2028.
\newblock \url{https://www.statista.com/statistics/591181/global-augmented-virtual-reality-market-size/}.   (2021).
\newblock
\newblock
\shownote{(Accessed 30.01.2023).}


\bibitem{issaSocialIsolationPsychological2021}
{H. Issa} {and} {E. Jaleel}. 2021.
\newblock \showarticletitle{Social Isolation and Psychological Wellbeing: Lessons from {{Covid-19}}}.
\newblock {\em Management Science Letters\/} {11}, 2 (2021), 609--618.
\newblock


\bibitem{jamesalexandercrewdsonEffectLonelinessElderly2016}
{{James Alexander Crewdson}}. 2016.
\newblock \showarticletitle{The {{Effect}} of {{Loneliness}} in the {{Elderly Population}}: {{A Review}}}.
\newblock {\em Healthy Aging \& Clinical Care in the Elderly\/}  {8} (March 2016), 1--8.
\newblock
\showISSN{11790601}
\showDOI{%
\url{http://dx.doi.org/10.4137/HACCE.S35890}}


\bibitem{kocaballiPersonalizationConversationalAgents2019}
{Ahmet~Baki Kocaballi}, {Shlomo Berkovsky}, {Juan~C. Quiroz}, {Liliana Laranjo}, {Huong~Ly Tong}, {Dana Rezazadegan}, {Agustina Briatore}, {and} {Enrico Coiera}. 2019.
\newblock \showarticletitle{The {{Personalization}} of {{Conversational Agents}} in {{Health Care}}: {{Systematic Review}}}.
\newblock {\em Journal of Medical Internet Research\/} {21}, 11 (Nov. 2019), e15360.
\newblock
\showDOI{%
\url{http://dx.doi.org/10.2196/15360}}


\bibitem{kocurFlexingMusclesVirtual2020a}
{Martin Kocur}, {Melanie Kloss}, {Valentin Schwind}, {Christian Wolff}, {and} {Niels Henze}. 2020.
\newblock \showarticletitle{Flexing {{Muscles}} in {{Virtual Reality}}: {{Effects}} of {{Avatars}}' {{Muscular Appearance}} on {{Physical Performance}}}. In {\em Proceedings of the {{Annual Symposium}} on {{Computer-Human Interaction}} in {{Play}}} {\em ({{CHI PLAY}} '20)}. {Association for Computing Machinery}, {New York, NY, USA}, 193--205.
\newblock
\showISBNx{978-1-4503-8074-4}
\showDOI{%
\url{http://dx.doi.org/10.1145/3410404.3414261}}


\bibitem{levineNonexerciseActivityThermogenesis2002}
{James~A. Levine}. 2002.
\newblock \showarticletitle{Non-Exercise Activity Thermogenesis ({{NEAT}})}.
\newblock {\em Best Practice \& Research Clinical Endocrinology \& Metabolism\/} {16}, 4 (Dec. 2002), 679--702.
\newblock
\showISSN{1521-690X}
\showDOI{%
\url{http://dx.doi.org/10.1053/beem.2002.0227}}


\bibitem{mansonEscalatingPandemicsObesity2004}
{JoAnn~E. Manson}, {Patrick~J. Skerrett}, {Philip Greenland}, {and} {Theodore~B. VanItallie}. 2004.
\newblock \showarticletitle{The {{Escalating Pandemics}} of {{Obesity}} and {{Sedentary Lifestyle}}: {{A Call}} to {{Action}} for {{Clinicians}}}.
\newblock {\em Archives of Internal Medicine\/} {164}, 3 (Feb. 2004), 249--258.
\newblock
\showISSN{0003-9926}
\showDOI{%
\url{http://dx.doi.org/10.1001/archinte.164.3.249}}


\bibitem{mentisCraftingImageSurgical2016}
{Helena~M. Mentis}, {Ahmed Rahim}, {and} {Pierre Theodore}. 2016.
\newblock \showarticletitle{Crafting the {{Image}} in {{Surgical Telemedicine}}}. In {\em Proceedings of the 19th {{ACM Conference}} on {{Computer-Supported Cooperative Work}} \& {{Social Computing}}} {\em ({{CSCW}} '16)}. {Association for Computing Machinery}, {New York, NY, USA}, 744--755.
\newblock
\showISBNx{978-1-4503-3592-8}
\showDOI{%
\url{http://dx.doi.org/10.1145/2818048.2819978}}


\bibitem{nooneLonelyTouchNarrative2022}
{Catrin Noone} {and} {Phoebe~E. {McKenna-Plumley}}. 2022.
\newblock \showarticletitle{Lonely for {{Touch}}? {{A Narrative Review}} on the {{Role}} of {{Touch}} in {{Loneliness}}}.
\newblock {\em Behaviour Change\/} {39}, 3 (Sept. 2022), 157--167.
\newblock
\showISSN{0813-4839, 2049-7768}
\showDOI{%
\url{http://dx.doi.org/10.1017/bec.2022.12}}


\bibitem{OnlineShoulderRehab}
{physiorehab}. 2010.
\newblock {Physio Fitness | Physio REHAB | Tim Keeley}.
\newblock \url{https://www.youtube.com/@physiorehab}.   (2010).
\newblock
\newblock
\shownote{(Accessed 30.01.2023).}


\bibitem{OnlineFitnessMarket}
{Rachita R}. 2020.
\newblock Online {{Fitness Market Statistics}} | {{Virtual Fitness Industry Forecast-}} 2027.
\newblock \url{https://www.alliedmarketresearch.com/virtual-online-fitness-market}.   (2020).
\newblock
\newblock
\shownote{(Accessed 08.06.2022).}


\bibitem{rodriguez-ayllonRolePhysicalActivity2019}
{Mar{\'i}a {Rodriguez-Ayllon}}, {Cristina {Cadenas-S{\'a}nchez}}, {Fernando {Est{\'e}vez-L{\'o}pez}}, {Nicolas~E. Mu{\~n}oz}, {Jose {Mora-Gonzalez}}, {Jairo~H. Migueles}, {Pablo {Molina-Garc{\'i}a}}, {Hanna Henriksson}, {Alejandra {Mena-Molina}}, {Vicente {Mart{\'i}nez-Vizca{\'i}no}}, {Andr{\'e}s Catena}, {Marie L{\"o}f}, {Kirk~I. Erickson}, {David~R. Lubans}, {Francisco~B. Ortega}, {and} {Irene {Esteban-Cornejo}}. 2019.
\newblock \showarticletitle{Role of {{Physical Activity}} and {{Sedentary Behavior}} in the {{Mental Health}} of {{Preschoolers}}, {{Children}} and {{Adolescents}}: {{A Systematic Review}} and {{Meta-Analysis}}}.
\newblock {\em Sports Medicine\/} {49}, 9 (Sept. 2019), 1383--1410.
\newblock
\showISSN{1179-2035}
\showDOI{%
\url{http://dx.doi.org/10.1007/s40279-019-01099-5}}


\bibitem{sayinkasarLifeLockdownSocial2021}
{Kadriye Sayin~Kasar} {and} {Emine Karaman}. 2021.
\newblock \showarticletitle{Life in Lockdown: {{Social}} Isolation, Loneliness and Quality of Life in the Elderly during the {{COVID-19}} Pandemic: {{A}} Scoping Review}.
\newblock {\em Geriatric Nursing\/} {42}, 5 (Sept. 2021), 1222--1229.
\newblock
\showISSN{0197-4572}
\showDOI{%
\url{http://dx.doi.org/10.1016/j.gerinurse.2021.03.010}}


\bibitem{schonauerMultimodalMotionGuidance2012b}
{Christian Sch{\"o}nauer}, {Kenichiro Fukushi}, {Alex Olwal}, {Hannes Kaufmann}, {and} {Ramesh Raskar}. 2012.
\newblock \showarticletitle{Multimodal Motion Guidance: Techniques for Adaptive and Dynamic Feedback}. In {\em Proceedings of the 14th {{ACM}} International Conference on {{Multimodal}} Interaction} {\em ({{ICMI}} '12)}. {Association for Computing Machinery}, {New York, NY, USA}, 133--140.
\newblock
\showISBNx{978-1-4503-1467-1}
\showDOI{%
\url{http://dx.doi.org/10.1145/2388676.2388706}}


\bibitem{spohrFakeNewsIdeological2017}
{Dominic Spohr}. 2017.
\newblock \showarticletitle{Fake News and Ideological Polarization: {{Filter}} Bubbles and Selective Exposure on Social Media}.
\newblock {\em Business Information Review\/} {34}, 3 (Sept. 2017), 150--160.
\newblock
\showISSN{0266-3821}
\showDOI{%
\url{http://dx.doi.org/10.1177/0266382117722446}}


\bibitem{PhysicalTherapyVirtual}
{The~EDGE Sports} {and} {Fitness}. 2022.
\newblock Physical {{Therapy Virtual}}.
\newblock \url{https://edgevt.com/physical-therapy-virtual/}.   (2022).
\newblock
\newblock
\shownote{(Accessed 30.01.2023).}


\bibitem{taylorSocialIsolationDepression2018}
{Harry~Owen Taylor}, {Robert~Joseph Taylor}, {Ann~W. Nguyen}, {and} {Linda Chatters}. 2018.
\newblock \showarticletitle{Social {{Isolation}}, {{Depression}}, and {{Psychological Distress Among Older Adults}}}.
\newblock {\em Journal of Aging and Health\/} {30}, 2 (Feb. 2018), 229--246.
\newblock
\showISSN{0898-2643}
\showDOI{%
\url{http://dx.doi.org/10.1177/0898264316673511}}


\bibitem{thompsoncoonDoesParticipatingPhysical2011}
{J. Thompson~Coon}, {K. Boddy}, {K. Stein}, {R. Whear}, {J. Barton}, {and} {M.~H. Depledge}. 2011.
\newblock \showarticletitle{Does {{Participating}} in {{Physical Activity}} in {{Outdoor Natural Environments Have}} a {{Greater Effect}} on {{Physical}} and {{Mental Wellbeing}} than {{Physical Activity Indoors}}? {{A Systematic Review}}}.
\newblock {\em Environmental Science \& Technology\/} {45}, 5 (March 2011), 1761--1772.
\newblock
\showISSN{0013-936X, 1520-5851}
\showDOI{%
\url{http://dx.doi.org/10.1021/es102947t}}


\end{thebibliography}

\end{document}